# ConTraNet: A single end-to-end hybrid network for EEG-based and EMG-based human machine interfaces

Omair Ali[1,4], Muhammad Saif-ur-Rehman[3], Tobias Glasmachers[2], Ioannis Iossifidis[3] and Christian Klaes[1]

[1] Faculty of Medicine, Department of Neurosurgery, University hospital Knappschaftskrankenhaus Bochum GmbH, Germany, [2]Institut für Neuroinformatik, Ruhr University Bochum, Germany, [3] Department of Computer Science, Ruhr-West University of Applied Science, Mülheim an der Ruhr, Germany; [4]Department of Electrical Engineering and Information Technology, Ruhr-University Bochum

## Abstract

**Objective:** Electroencephalography (EEG) and electromyography (EMG) are two non-invasive bio-signals, which are widely used in human machine interface (HMI) technologies (EEG-HMI and EMG-HMI paradigm) for the rehabilitation of physically disabled people. Successful decoding of EEG and EMG signals into respective control command is a pivotal step in the rehabilitation process. Recently, several Convolutional neural networks (CNNs) based architectures are proposed that directly map the raw time-series (EEG and EMG signal) into decision space (intended action of the user). Since CNNs are end-to-end learning algorithms, the process of meaningful features extraction and classification are performed simultaneously. However, these networks are tailored to the learn the expected characteristics of the given bio-signal. Henceforth, the implication of these algorithms is usually limited to single HMI paradigm. In this work, we addressed the question that can we build a single architecture which is capable of learning distinct features from different HMI paradigms and still successfully classify them. **Approach:** In this work, we introduce a single hybrid model called ConTraNet, which is based on CNN and Transformer architectures that is equally useful for EEG-HMI and EMG-HMI paradigms. ConTraNet uses CNN block to introduce inductive bias in the model and learn local dependencies, whereas the Transformer block uses the self-attention mechanism to learn the long-range or global dependencies in the signal, which are crucial for the classification of EEG and EMG signals. **Main results:** We evaluated and compared the ConTraNet with state-of-the-art methods on three publicly available datasets (BCI Competition IV dataset 2b, Physionet MI-EEG dataset, Mendeley sEMG dataset) which belong to EEG-HMI and EMG-HMI paradigms. ConTraNet outperformed its counterparts in all the different category tasks (2-class, 3-class, 4-class, and 10-class decoding tasks). **Significance:** Most HMI studies introduce the algorithms that are tailored to the characteristics of its expected bio-signal and validate their results on the dataset/s, which belong to only single paradigm. Contrarily, we introduced ConTraNet and validated the results on two different HMI paradigms, which contain the data of 2, 3, 4 and 10-classes. Furthermore, the generalization quality of ConTraNet remains equally good for both paradigms, which suggest that ConTraNet is robust to learn distinct features from different HMI paradigms and generalizes well as compared to the current state of the art algorithms.

# 1 Introduction and Related Works

Human machine interface (HMI) is a technology that provides a connection and control between user and the external machine. The connection and control offered by HMIs play a vital role in medical applications such as neural rehabilitation of physically disabled people and enabling the amputees to control upper limb prostheses by establishing an alternate pathway between human and the device [1]. Two widely used HMI paradigms in medical applications or rehabilitation engineering are brain computer interface (BCI) [2], [3] and electromyography-based HMI (EMG-HMI) [4], [5]. In BCI paradigms, brain signals depicting the user's intent are translated into corresponding command signals to control an external device [6], whereas in EMG-HMI paradigm, muscle movements are employed to channel the output device [7].

Both paradigms generally consist of five processing steps [5], [8]: data recording, pre-processing, feature extraction, inference, and control. The first step as mentioned earlier is to record the data. In both archetypes, the data is recorded either by invasive or non-invasive methods [9]–[14]. Invasive methods in BCI require implanting electrodes in brain at areas of interest, whereas electroencephalography signals (EEG) are widely used in non-invasive BCI systems. EEG signals are the electrical brain activity recorded from electrodes which are placed on the scalp. Similarly, needle electrodes are planted inside the muscle in invasive EMG-HMI systems, whereas surface EMG (sEMG) signals are one of the most common used signals in non-invasive EMG-HMI systems. sEMG signals are the electrical activity generated during the muscle contraction which is recorded using electrodes placed on the skin. In the second step, the recorded data is digitized and preprocessed. After the second stage, the features are extracted from the preprocessed data which are then employed by the decoder in the fourth step to translate the neural or muscle activity into corresponding user intention which is then finally transformed into a control signal in the last step to control an output device.

Non-invasive BCI and EMG-HMI systems have gained traction in medical applications because of their convenient, less risk prone and non-complex nature [15]–[19]. In EEG based non-invasive BCI systems, movement related signals obtained from the motor cortex area of the brain are called motor execution (ME), if an actual limb movement is performed, or motor imagery (MI) if no overt motion of the limb is performed. The focus of this study is on decoding the movement related information and muscle activity based on MI-EEG and sEMG signals respectively.

Conventionally, the successful mapping of MI-EEG and sEMG signals into respective control commands follow two steps protocol. In the first step, meaningful features are extracted from the signal which are then employed by the classifier in the next step to output the corresponding user intention. There are various methods that are generally employed to extract features and perform inference [20]–[22]. In [21], MI-EEG signal is first decomposed into intrinsic mode functions using multivariate empirical mode decomposition method. The features are then extracted using short time Fourier transform (STFT) from the most significant mode which are later used by k-Nearest Neighbor (kNN) for classification. In [23], anchored-STFT is proposed to transform the time domain MI-EEG signal into image representation in frequency domain which is then fed to Convolutional Neural Network (CNN) architecture named Skip-Net for feature extraction and classification. In [22], Filter bank common spatial pattern (FBCSP) is introduced to extract the discriminative patterns from the MI-EEG signals. The extracted features are then classified using various machine learning algorithms such as Naïve Bayesian Panzer Window (NBPW), Fisher Linear Discriminant (FLD), Support Vector Machine (SVM), Classification and Regression Tree (CART) and, kNN. In [24] Morlet wavelet is used to extract rhythmic information from mu band of the MI-EEG signal which is then classified using Bayes theorem. In [25] wavelet coefficients and autoregressive parameter model are used to extract the features from MI-EEG signal and Linear Discriminant Analysis (LDA) is used as a classifier. Similarly, in [26] Discrete Wavelet Transform (DWT) is used to obtain features from sEMG signals which are then classified using SVM. In [27] time-domain and frequency domain features are extracted from sEMG signals and classified using SVM using radial basis function. In [28] Logarithmic Spectrogram-based Graph Signal (LSGS) is employed to extract meaningful features from sEMG signals which are then classified using a hybrid model named as AdaBoost k-means (AB-k-means).

Despite the successful utility of conventional methods to translate MI-EEG and sEMG signals to respective control commands, they pose a potent drawback in their methodology, which is the hand engineered feature extraction process. To alleviate this problem, now a days, Deep Learning algorithms are deployed in several studies which follow end-to-end learning protocol, which in turn bypass the need to select meaningful features manually [29]–[31]. In [29], a CNN architecture named EEGNet is introduced for EEG based BCI paradigms. It automatically extracts and classify the features from the raw EEG signals in end-to-end learning fashion. In [30], different CNN architectures namely DeepConvNet, ShallowConvNet (ShallowNet in rest of the manuscript), HybridConvNet and, ResidualConvNet are presented to perform end-to-end MI-EEG decoding. In [31], a CNN architecture with hybrid convolution scale namely HS-CNN is introduced for the classification of MI-EEG signals. Here the raw signals are first bandpass filtered in user specified three frequency ranges which are then used for automatic feature extraction and classification by multi scale convolution kernels. Henceforth, we consider it as semi end-to-end learning method.

Although CNN based architectures have shown leading results in the classification of MI-EEG and sEMG signals as well as in other domains of BCI [32], [33], they still present a drawback in their architecture, that is: the feature extractors based on CNN architecture considerably rely on the selection of kernel sizes [31], [34], [35]. Large kernel size obstructs its exploration of deep feature whereas, the small kernel size dampers the receptive field [36] which restricts it from learning the global context that is usually essential for the classification of MI-EEG and sEMG signals. Henceforth to capture the wide range of internal relationship between different time stamps of MI-EEG and sEMG signals or global context, a deep architecture is required which significantly increases the computational cost. Secondly, since each convolutional layer of a CNN architecture shares the weight which means that it uniformly attends to each part of the MI-EEG and sEMG signal in that layer to learn meaningful patterns. However, MI-EEG and sEMG signals are non-stationary in nature henceforth, each part of the signal might not be equally significant for correctly classifying them. Therefore, more attention should be paid to the relevant parts of the signal compared to less relevant parts. Henceforth, to learn long-distance features and to attend specific parts of the signals, some studies have employed weighted attention mechanism in their architectures [37]–[40]. In [37], an attention based Bi-Long Short Term Memory (Bi-LSTM) method is introduced for visual object classification using EEG. Here the forget gates of LSTM are replaced with attention gate. In

[38], an attention based convolutional recurrent neural network (ACRNN) is proposed to extract more discriminative features from EEG signals for emotion recognition. In [39] and [40], a convolutional recurrent attention model (CRAM) and graph based CRAM (G-CRAM) is introduced respectively to extract high level and discriminative features from motor imagery signals.

Even though attention-based LSTM and RNN have gained success in some scenarios, yet they pose a limit in learning long-distance context dependencies in sequence-to-sequence modeling, which is critical for successful decoding of MI-EEG and sEMG signals and secondly, they lack the capabilities to parallelize the processing of the signal because of their recursive nature. To overcome these limitations, a new architecture based on self-attention mechanism called, the Transformer is introduced in [41]. The transformer architecture is originally introduced for natural language processing, but it has found its way to many other applications due to its effectiveness in learning long-range global dependencies and ability to parallelize the computation [36], [42]. However, its exploration in classifying MI-EEG and sEMG signals remains limited. It is because, transformers lack the inductive bias in their design. It must be learned from scratch which usually require a very large amount of data to pretrain them and then transfer to mid-sized recognition tasks [42]. Since the data scarcity issue prevails in EEG and sEMG based HMI paradigms, the use of transformers in this field also remains limited. In [43], a model called Spatial-Temporal Tiny Transformer (S3T) is presented for EEG decoding. Here EEG signal is first filtered using the idea of Common Spatial Pattern (CSP), and then the outputs of multiple one-versus-rest (OVRs) are stacked to form the input for S3T. Like conventional approaches, this method also follows the two-step protocol for the classification the of EEG signals, hand engineered feature extraction and then classification.

Unlike transformer architectures, CNN based models have built in hard inductive bias in their design which is very useful for smaller datasets. On the other hand, CNN architectures need big size kernels to learn long-range dependencies which in turn restricts learning deep features. On the contrary, transformer architectures use self-attention mechanism which learns the global dependencies in a sequence effectively without compromising on computational and statistical efficiency. To address these issues, in this study, we introduced a hybrid model that combines the strengths of both CNN and transformer architectures to introduce inductive bias and the ability to learn long-range global dependencies in the sequence. We named the hybrid model 'ConTraNet'. The experimental evaluation during this study established that ConTraNet outperformed its counterparts on three publicly available MI-EEG and sEMG datasets.

# 2 Materials and Methods

## 2.1 Data Description

In this study, the classification of physiological signals is performed. Two types of physiological signals are used for the classification which are, MI-EEG signals and sEMG signals. In case of MI-EEG signals, the data comes from two datasets which are BCI Competition IV dataset 2b and Physionet MI-EEG dataset. However, in case of sEMG signals, the data comes from Mendeley Data – sEMG dataset.

### 2.1.1 BCI Competition IV dataset 2b (MI-EEG)

The dataset consists of two classes of MI-EEG signals of the left (L) and right hand (R) of 9 subjects. Each subject performed five sessions (01T, 02T, 03T, 04E and 05E). The MI-EEG data is recorded using three bipolar electrodes (C3, Cz and C4), sampled at 250 Hz and bandpass filtered between 0.5 Hz and 100 Hz, and a notch filter at 50 Hz is applied. For training and testing, the MI-EEG signal from second 3 to second 5.5 (2.5s in total) is considered for each trial. The timing window used here is the same as used in [23]. We perform 2-class (L/R) classification on this dataset. The detailed description of the dataset is available in [44].

### 2.1.2 Physionet MI-EEG dataset

The dataset contains the MI/ME-EEG recordings of 109 subjects. Four subjects are dropped from the dataset since they contained a varied number of trials, resulting in 105 subjects to be used in this study. Subjects performed both ME and MI-tasks, but however, we just focused on the classification of MI-tasks in this work. MI-EEG signals are recorded with the BCI-2000 system using 64 electrodes, sampled at 160 Hz. Each subject performed three runs of MI-tasks of left fist (L) or right fist (R). Additionally, they also performed three runs of MI-tasks of both fists (B) or both feet (F). Moreover, the baseline run provides the resting data (0) where the subjects did not receive any cues. Each run is 120s long and contains 14 MI-trials, resulting in 21 trials per class per subject. For this dataset, we distinguish between 2-class (L/R), 3-class (L/R/0) and 4-class (L/R/0/F) MI-tasks. The detailed description of the dataset is available in [45].

### 2.1.3 Mendeley Data - sEMG

The dataset contains the 4-channel sEMG recordings of 10 different hand gestures from 40 subjects. The data are recorded with a BIOPAC MP36 device using Ag/AgCl surface bipolar electrodes, sampled at 2 kHz and bandpass filtered between 5 Hz and 500 Hz using a sixth-order Butterworth bandpass filter and a notch filter is applied at 50 Hz. Each subject participated in five repetitive cycles with 30s rest in between cycles. All cycles contain the 10 hand gestures. Each cycle lasts for 104s, where each hand gesture is 6s long with 4s resting in between two hand gestures. The hand gestures performed by the subjects are rest or neutral state, extension of the wrist, flexion of the wrist, ulnar deviation of the wrist, radial deviation of the wrist, grip, abduction of all fingers, adduction of all fingers, supination, and pronation. For this dataset, we distinguish between 10-class sEMG signals. The detailed explanation and description of the data is presented in [46].

## 2.2 Methods

Here, we introduce ConTraNet, an end-to-end CNN-transformer based hybrid architecture, that can be applied across different physiological signals based human machine interaction (HMI) paradigms such as brain computer interface (BCI) etc. Before we describe ConTraNet in detail, we explain about CNN and Transformer architectures.

### 2.2.1 Convolutional Neural Network (CNN)

CNN, as the name suggests, consists of convolutional layers as its backbone. The layers parameters focus on the use of learnable kernels which are essential to automatically extract the meaningful features from the raw inputs. The process of feature extraction takes place in several stages by stacking the multiple convolutional layers. Convolutional layer first learns to map the raw input into low-level features, which are then mapped to mid-level features, finally, mid -level features are mapped to high-level feature. High-level representation of the input suppresses the redundant information and highlight the distinguishable pivotal information.

In addition to the convolutional layer as the main building block of a CNN architecture, it may also contain few other layers [47] such as normalization layer which ensures that the gradients while training remain in a certain range, activation layer which adds the non-linearity to the output of a convolutional layer, pooling layer which reduces the dimension of the feature maps and extracts the most relevant features from the output of convolutional layer, regularization layer to avoid overfitting and fully connected layer which is analogous to standard artificial neural networks (ANNs) in its role.

Similarly, ConTraNet also consists of a CNN block with convolutional layer, normalization layer, activation layer, and pooling layer.

CNNs have produced many state-of-the-art results in solving computer vision tasks such as image classification, object detection and image segmentation etc. One reason behind this success is the translation invariance or equivariance of CNNs. Secondly, the ability of the CNNs to learn deep features by using small size kernels also play a vital role in many patterns' recognition tasks. However, at the same time, the small size kernels limit its ability to learn long rage dependencies which is usually vital in time series signals such as MI-EEG and sEMG signals. This issue is generally addressed by increasing the receptive field of the architecture by stacking more convolutional layers, hence deepening the model. Increasing the depth of the model also enhances the trainable parameters which in result increases the computational cost and the chances of overfitting. Overfitting is a curse that could badly hamper the generalization ability of the learned model and hence be avoided at all costs.

### 2.2.2 Transformers

Transformers are originally introduced for machine translation in [41] and have since become the de facto method for many natural language processing (NLP) tasks. The network architecture mainly relies on attention mechanism and disregards the recurrence and convolutions completely. The sole reliance on attention gives it the ability to draw global dependencies between the input and the output. The standard transformer architecture as shown in **Figure 1** consists of an encoder and a decoder block. Transformer-encoder module encodes the input, which is 1D sequence of token embeddings, to a sequence of continuous representations. Whereas the Transformer-decoder generates the output sequence one element at a time provided the encoded representation from encoder.

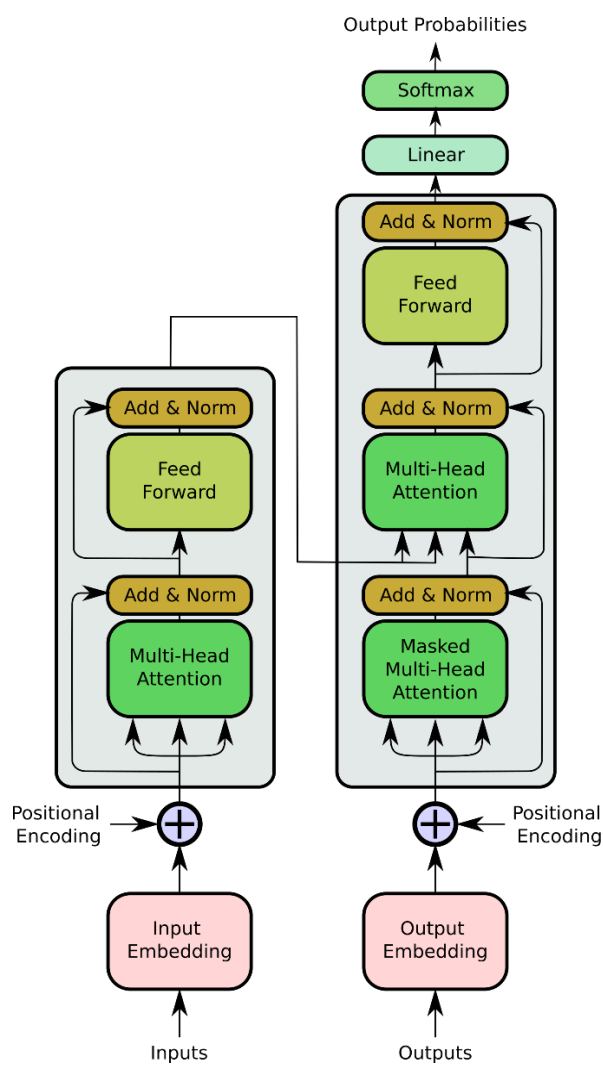

*Figure 1: Transformer model architecture.*

**Encoder block:** Transformer-encoder consists of several identical layers called encoder-layers. Each encoder-layer has two sub-layers namely multi-head self-attention layer and position wise fully connected feed-forward layer. Each sub-layer is followed by layer normalization [48]. The input to each sublayer is connected to the followed normalization layer via residual connection [49].

**Decoder block:** Like encoder, decoder also consists of several identical layers called decoder-layers. Each decoder layer has three sub-layers. Two of them are multi-head self-attention layers and the remaining is the fully connected feed-forward layer. One of the multi-head attention layers attends to the encoded representation of the encoder block, whereas the other multi-head attention layer has masking mechanism which prevents the decoder positions to attend to subsequent positions which ensures that the output for position $i$ can only depend on the outputs of positions less than position $i$. Similar to encoder block, each sub-layer is followed by layer normalization and the connection between the input to each sub-layer and the following normalization layer is established through the residual connection.

**Self-attention:** The attention as shown in **Figure 2** maps the query and a set of key-value pairs to an output. The query, key and value vectors are obtained through the linear projection of input vector using three linear layers. The attention as shown in **equation 1** is then performed by computing the dot products of query with all the keys, which are then scaled by the dimension of the model which is same as the dimension of query and keys. The softmax function is then applied to the result to obtain the weights on values. The resultant is finally multiplied by the value vector to obtain the complete attention score.

$$\text{Attention}(Q, K, V) = \text{softmax}\left(\frac{QK^T}{\sqrt{d_k}}\right)V \qquad (1)$$

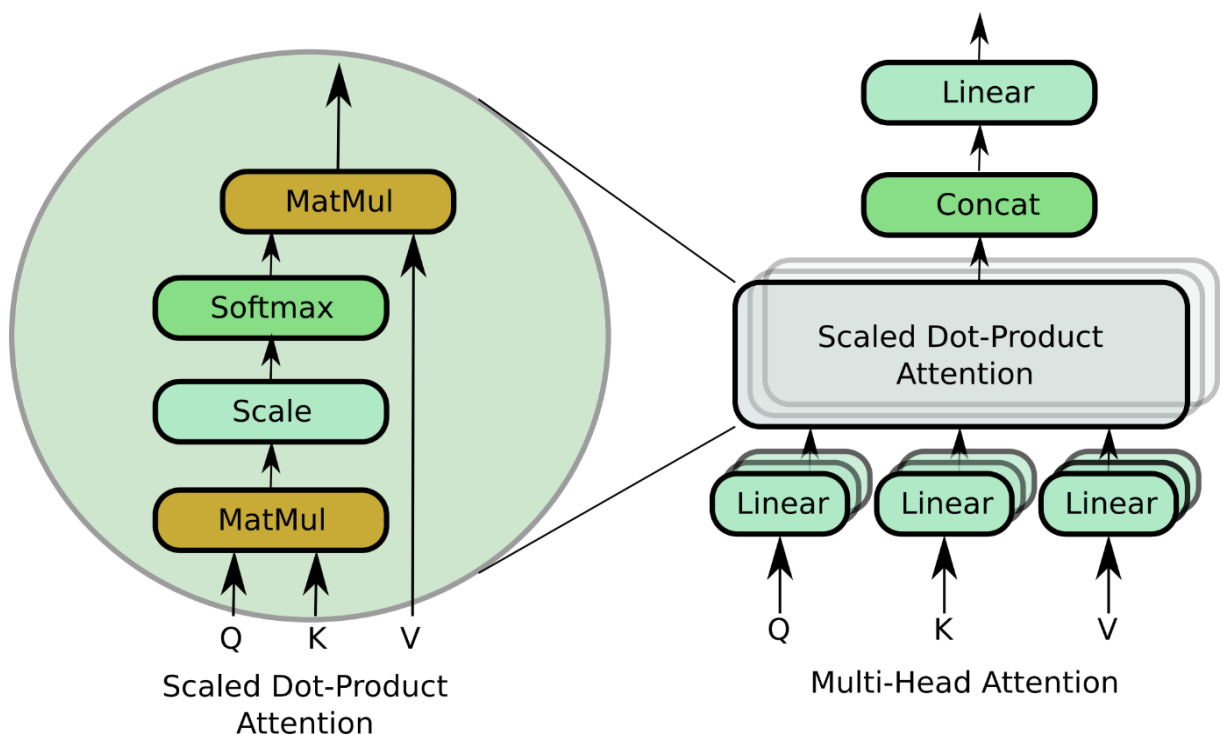

*Figure 2: Schematic diagram of Scaled Dot-Product Attention (left) and Multi-Head Attention (right).*

**Multi-Head attention:** Multi-head attention layer as shown in **Figure 2** consists of multiple self-attention mechanisms that run in parallel and are referred as heads. Here the queries, keys and values are linearly projected $h$ times with different linear projections. Self-attention function is then performed on each of the projected versions of queries, keys, and values in parallel which are concatenated and once again projected to output the result. Multi-Head attention can be represented as **equation 2**.

$$\text{Multi-Head}(Q, K, V) = \text{Concatenate}(\text{head}_1, \ldots, \text{head}_h)W^O \qquad (2)$$

$$\text{where head}_i = \text{Attention}(QW_i^Q, KW_i^K, VW_i^V)$$

Here $W_i^Q \in \mathbb{R}^{d_{model} \times d_k}$, $W_i^K \in \mathbb{R}^{d_{model} \times d_k}$, $W_i^V \in \mathbb{R}^{d_{model} \times d_v}$ and $W^O \in \mathbb{R}^{hd_v \times d_{model}}$ are the projection matrices. $d_{model}$ is the dimension of the model and $h$ represents number of heads.

**Embeddings and Positional Encoding:** Embeddings are the learned representations of input and output tokens in form of a vector that meets the dimension criterion of the model. Transformer model is permutation invariant because it does not contain any recurrence and convolution operations. To add some information about the position of the tokens in the sequence, positional encodings are added to the input embeddings in both encoder and decoder blocks.

### 2.2.3 ConTraNet

ConTraNet as shown in **Figure 3** is a hybrid architecture which consists of CNN and the Transformer block.

**CNN block:** CNN block induces inductive biases such as translation equivariance and locality in the architecture which is significant in small-size datasets. Additionally, it also learns or extracts the temporal features from the signals. Apart from this, it also learns the local dependencies within the kernel length which are important for online decoding. Another perspective of CNN block is that it guides the Transformer block since it extracts the relevant local temporal features and presents them to the Transformer block to find the global dependencies among them.

The CNN block in ConTraNet contains a convolutional layer, pooling layer, regularization layer, a normalization layer, and an activation layer. The convolutional layer uses 2D convolutional filters of size (1, 1/sampling frequency) to acquire the frequency resolution of 2 Hz. Henceforth it allows the model to capture frequency information at 2 Hz and above. Batch normalization is then applied to the feature maps followed by exponential linear unit (ELU) non-linearity [50]. Average pooling layer of size (1,8) is used to reduce the sampling rate of the signal to 1/8$^{th}$ of the 1/sampling frequency. To regularize the model and prevent it from overfitting, dropout technique [51] with dropout rate of 0.5 is used in CNN block. Spatial dropout is used for MI-EEG signals whereas normal dropout is used for sEMG signals. Additionally, each temporal filter is regularized by using a maximum norm constraint of 0.25 on its weights.

**Transformer block:** Transformer block learns the global or long-range dependencies, which play a vital role in successful decoding of MI-EEG and sEMG signals. Additionally, it learns to pay more attention to the relevant parts of the signals which are vital for the correct classification while suppressing the non-relevant parts.

The Transformer block in ConTraNet contains only the encoder followed by classification head (multi-layer perceptron (MLP) head and a softmax layer). The output of the CNN block is the input of the Transformer block. Since the Transformer model is originally presented for NLP tasks and only accepts the embeddings of the sequence tokens as its inputs, henceforth could not be directly used for images. However [42] presented an approach to transform an image into sequence of linear embeddings which could be used as inputs to the Transformer model. Inspired by this approach, the feature maps of CNN block are transformed into respective embeddings which are provided as input to the Transformer block.

The CNN block outputs the feature maps of shape [$N_{ch}$ x $N_s$ x $N_f$], where $N_{ch}$ = number of channels (EEG or EMG electrodes), $N_s$ = number of samples and, $N_f$ = number of filters. Firstly, image patches of shape [$N_{ch}$ x 2 x $N_f$] are extracted from the feature maps. The image patches are overlapped by factor of $N_s$ = 1. The extracted patches are linearly projected to obtain the learnable patch embeddings which are then added with learnable positional encodings to include the position information. The embedded patches (patch embeddings + positional embeddings) are sent to the Transformer Encoder as its input.

The Transformer encoder in ConTraNet consists of a single encoder layer. The encoder layer contains multi-headed self-attention and a MLP block. The MLP block contains two hidden layers with an ELU non-linearity. Layer normalization is applied before MLP and attention blocks, whereas residual connection is applied after MLP and attention blocks. The Transformer encoder is followed by a classification head which is implemented by another MLP with one hidden layer and a softmax layer. Dropout of 0.7 is used in MLP block of encoder layer as well as classification head to regularize the model and prevent overfitting. Additionally, each hidden layer is regularized by using a maximum norm constraint of 0.25 on its weights.

Since the Transformer uses constant latent vector size $d_{model}$, also referred as model dimension, through all its layers, henceforth, all the transformations (patch embeddings, positional embeddings, output of multi-head self-attention, output of MLP block in encoder layer) applied match the model dimension.

The hyperparameters used for the training of ConTraNet are shown in **Table 1**.

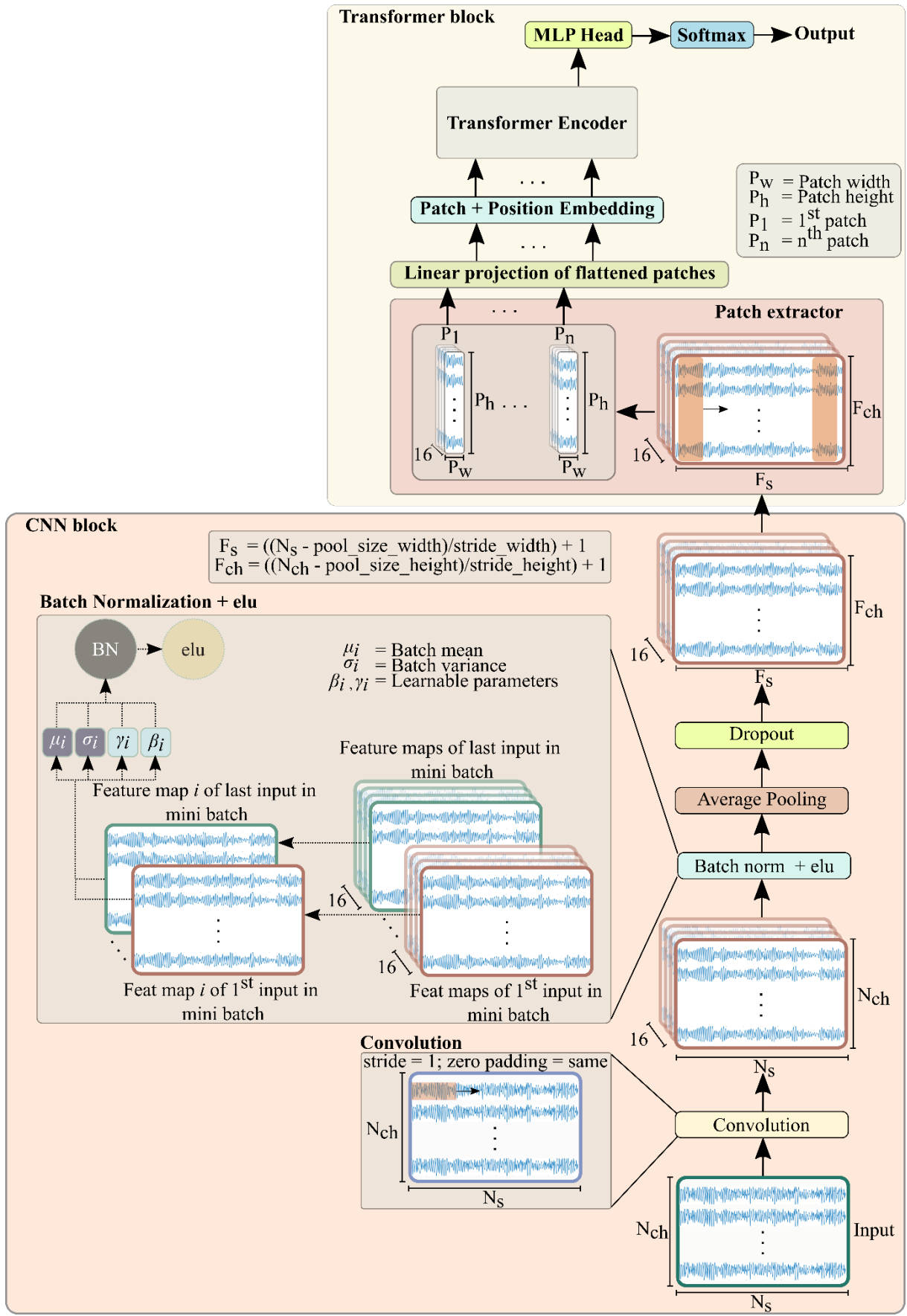

*Figure 3:* ConTraNet. The model architecture.

*Table 1: Hyperparameters of ConTraNet.*

| Hyperparameters | Values |
|---|---|
| Optimization algorithm | Adam |
| Epochs | 100 |
| Learning rate | 0.001, 0.0001 at epochs 0 and 50 |
| No. of convolutional kernels | 16 |
| Padding in Convolutional layer | Same |
| $d_{model}$ | 32 |
| No. of heads | 8 |
| $P_w$, $P_h$ | 2, $F_{ch}$ |
| Stride in Patch extractor | 1 |
| Padding in Patch extractor | Valid |
| Hidden units in Encoder MLP block | $2*d_{model}$, $d_{model}$ |
| Hidden units in MLP Head | 112 |

# 3 Results

A comprehensive performance evaluation of the proposed method and its comparison with existing state-of-the-art methods is presented here. Three very popular and state-of-the-art methods, which follow the end-to-end learning protocol and are chosen as baseline methods to compare the performance with ConTraNet, are EEGNet [29], DeepConvNet [30] and ShallowNet [30]. Three versatile, publicly available benchmark datasets are used for performance comparison and validation of the results. The datasets are as follows.

- BCI Competition IV, dataset 2b
- Physionet MI-EEG dataset
- Mendeley Data – sEMG

No additional pre-processing or filtering is applied on the datasets other than what originally performed by the providers of the datasets. The detailed description of the datasets is presented in subsection 'Data description' of section 'Materials and Methods'.

## 3.1 Evaluation metrics and Statistical analysis

We used accuracy as the evaluation metrics and performed paired t-test to establish the statistical significance of the proposed method.

## 3.2 Evaluation protocol

To have a fair comparison between state-of-the-art methods and the proposed architecture, we performed K-fold cross validation on all datasets. The evaluation protocol for datasets is as follows.

**BCI Competition IV dataset 2b:** Since the dataset comprised of MI-EEG data of 9 subjects, we employed 9-fold cross validation (Global) resulting into 8 subjects for training and the remaining 1 subject for testing in each fold. Additionally, we performed subject specific decoding (SS), where we trained ConTraNet on training sessions (Sessions 01T, 02T, and 03T) and evaluated on test sessions (Sessions 04E and 05E). Moreover, we also evaluated the impact of transfer learning on the performance of ConTraNet for each subject individually, where we finetuned the global model of each fold obtained after cross validation, on the training sessions of the respective left out subject (Global to SS) and evaluated on the test sessions.

**Physionet MI-EEG dataset:** The dataset comprised of MI-EEG data of 109 subjects. However, 4 subjects are excluded due to imbalanced number of trials. Consequently, 105 subjects are used for performance evaluation. Here, we performed 5-fold cross validation, resulting into 84 subjects for training and the remaining 21 subjects for testing in each fold.

**Mendeley Data – sEMG:** The dataset includes EMG data of 40 subjects. Here, we performed 5-fold cross validation which implies 32 subjects for training and the remaining 8 for testing in each fold.

The summary of the datasets and the evaluation protocol used to evaluate and compare the performance of ConTraNet with state-of-the-art methods is presented in **Table 2.**

*Table 2: Summary of datasets and the evaluation protocol used in this study.*

| Dataset | Signal type | No. of Subjects | No. of Classes | Evaluation Protocol |
|---|---|---|---|---|
| **BCI Competition IV dataset 2b** | MI-EEG | 9 | 2 | 9-Fold CV (Global), Subject Specific decoding (SS) |
| **Physionet dataset** | MI-EEG | 105 | 2, 3, 4 | 5-Fold CV (Global) |
| **Mendeley dataset** | sEMG | 40 | 10 | 5-Fold CV (Global) |

## 3.3 Performance evaluation on BCI Competition IV dataset 2b

Here, we report the performance of ConTraNet on BCI Competition IV dataset 2b and its comparison with state-of-the-art methods. **Table 3** shows the 9-fold cross validation performance of each method. It is shown in **Table 3**, that ConTraNet outperformed all its counterparts. ConTraNet obtained 4.2% improvement in average classification performance compared to ShallowNet, whereas the improvement is 2.9% and 2.4% compared to DeepConvNet and EEGNet respectively.

*Table 3: Performance comparison of ConTraNet with state-of-the-art methods on Competition IV dataset 2b using 9-fold cross validation.*

|  | Folds | EEGNet (Global) | DeepConvNet (Global) | ShallowNet (Global) | ConTraNet (Global) |
|---|---|---|---|---|---|
| **9-Fold CV** **2-class** **L / R** | 1 | 70.69 | 73.33 | 70.56 | 72.92 |
|  | 2 | 68.24 | 69.26 | 66.32 | 72.94 |
|  | 3 | 64.03 | 63.06 | 62.78 | 63.75 |
|  | 4 | 74.49 | 73.78 | 75.95 | 83.51 |
|  | 5 | 82.43 | 80.68 | 76.89 | 82.70 |
|  | 6 | 78.75 | 74.72 | 71.25 | 80.69 |
|  | 7 | 80.09 | 81.81 | 80.03 | 84.44 |
|  | 8 | 76.84 | 75.13 | 73.82 | 77.37 |
|  | 9 | 72.08 | 70.97 | 73.06 | 70.83 |
|  | **Avg** | 74.18 | 73.63 | 72.29 | **76.57** |

**Table 4** shows the subject specific decoding performance of ConTraNet and its comparison with state-of-the-art methods. It also shows the effect of transfer learning on the performance of ConTraNet. It is seen from **Table 4**, that ConTraNet obtained the highest average accuracy of 85.29% which is 5.57% higher than DeepConvNet and 3.15% and 2.38% higher compared to ShallowNet and EEGNet respectively. However, finetuning of global models of each fold of ConTraNet on respective subject specific data, yielded the average classification accuracy of 86.98% which is 1.69% higher than average classification performance of subject specific models of ConTraNet. The improvement seconds the theoretical notion of transfer learning.

*Table 4: Performance comparison of ConTraNet with state-of-the-art methods on BCI Competition IV dataset 2b using subject specific decoding.*

|  | Sub | EEGNet (SS) | DeepConvNet (SS) | ShallowNet (SS) | ConTraNet (SS) | ConTraNet (Global to SS) |
|---|---|---|---|---|---|---|
| **Subject** **specific** **decoding** **2-class** **L / R** | 1 | 68.12 | 55.00 | 70.94 | 75.31 | 76.88 |
|  | 2 | 68.44 | 67.50 | 64.69 | 70.62 | 72.15 |
|  | 3 | 85.94 | 83.13 | 71.56 | 88.40 | 89.00 |
|  | 4 | 97.81 | 94.38 | 92.50 | 97.19 | 98.00 |
|  | 5 | 95.31 | 86.56 | 96.50 | 92.31 | 95.00 |
|  | 6 | 72.19 | 71.25 | 85.60 | 77.50 | 87.50 |
|  | 7 | 89.06 | 87.50 | 85.62 | 88.19 | 90.30 |
|  | 8 | 94.38 | 92.19 | 89.38 | 95.31 | 95.00 |
|  | 9 | 75.00 | 80.00 | 82.50 | 82.81 | 79.00 |
|  | **Avg** | 82.91 | 79.72 | 82.14 | **85.29** | **86.98** |

## 3.4 Performance evaluation on Physionet MI-EEG dataset

To further validate the performance of ConTraNet, we evaluated it's results on another large scale publicly available dataset. Here, we present the performance comparison of ConTraNet with state-of-the-art methods on Physionet MI-EEG dataset. Since the dataset comprised of left fist (L), right fist (R), both fists (B), both feet (F) MI-tasks as well as the resting data (0), henceforth, the performance evaluation is made on 2-class (L/R), 3-class (L/R/0), and 4-class (L/R/0/F) MI-tasks.

**Table 5** shows the performance comparison of ConTraNet with state-of-the-art methods on 2-class, 3-class, and 4-class MI-tasks on Physionet MI-EEG dataset. It is shown in **Table 5**, that ConTraNet outperforms its counterparts in classification of 2-class, 3-class, and 4-class MI-tasks using 5-fold cross validation.

In 2-class MI-EEG decoding, ConTraNet obtained the average classification accuracy of 83.61% which is 3.09% higher than ShallowNet, whereas it yielded an improvement of 2.67% and 1.8% compared to DeepConvNet and EEGNet respectively.

In case of 3-class MI-EEG classification, ConTraNet achieved 74.38% average decoding accuracy of 5-folds. Here, it showed an improvement of 2.97% in classification accuracy compared to ShallowNet. However, the improvement is 2.43% and 1.93% compared to EEGNet and DeepConvNet respectively.

In the event of classification of 4-class MI-EEG task, ConTraNet exhibited an average improvement of 4.2% against the state-of-the-art methods. It attained the average classification accuracy of 65.44% which implies an improvement of 6.03% compared to ShallowNet, whereas these numbers are 2.82% and 3.80% compared to DeepConvNet and EEGNet respectively.

*Table 5: Performance comparison of ConTraNet with state-of-the-art methods on 2-class (left fist against right fist) Physionet MI-EEG dataset using 5-fold cross validation.*

|  | Folds | EEGNet (Global) | DeepConvNet (Global) | ShallowNet (Global) | ConTraNet (Global) |
|---|---|---|---|---|---|
| **5-Fold CV** | 1 | 84.47 | 81.97 | 80.73 | 84.24 |
|  | 2 | 79.48 | 77.78 | 78.12 | 81.18 |
| **2-class** | 3 | 83.33 | 87.53 | 83.67 | 87.07 |
| **L / R** | 4 | 79.25 | 75.62 | 77.44 | 80.61 |
|  | 5 | 82.43 | 81.75 | 82.65 | 84.61 |
|  | **Avg** | 81.79 | 80.93 | 80.52 | **83.61** |
| **5-Fold CV** | 1 | 74.07 | 70.52 | 71.13 | 74.98 |
|  | 2 | 70.60 | 71.96 | 70.75 | 74.38 |
| **3-class** | 3 | 75.74 | 77.17 | 74.75 | 79.82 |
| **L / R / 0** | 4 | 69.92 | 71.05 | 69.77 | 70.22 |
|  | 5 | 69.39 | 71.50 | 70.60 | 72.49 |
|  | **Avg** | 71.94 | 72.44 | 71.40 | **74.38** |
| **5-Fold CV** | 1 | 65.42 | 63.32 | 59.86 | 67.23 |
|  | 2 | 60.49 | 65.31 | 58.73 | 64.51 |
| **4-class** | 3 | 65.25 | 66.61 | 62.19 | 69.27 |
| **L / R / 0 / F** | 4 | 56.25 | 57.48 | 56.86 | 61.05 |
|  | 5 | 60.49 | 60.38 | 59.41 | 65.14 |
|  | **Avg** | 61.63 | 62.62 | 59.41 | **65.44** |

## 3.5 Performance evaluation on Mendeley Data-sEMG

To further establish the significance of ConTraNet and its abilities to learn distinct features across different HMI paradigms, we also evaluated its performance on physiological dataset which is different than brain signals (MI-EEG signals). We opted the data of muscle activities, Mendeley Data-sEMG, for this purpose since it is a timeseries physiological data. Here, we describe the performance comparison of state-of-the-art methods with ConTraNet on the classification of 10-classes of Mendeley Data-sEMG using 5-fold cross validation. Contrary to using an overlapping sliding window approach (approx. 200ms) to decode the EMG signals in an online fashion, we deployed the models in offline mode to decode the complete trial at once. The decoding performance of ConTraNet and the existing state-of-the-art methods is presented in **Table 6**. It is shown in **Table 6** that, ConTraNet surpassed its counterparts by manifesting an average improvement of 8.6% on average classification accuracy of 10-class EMG signals.

ConTraNet acquired an average classification accuracy of 77.15% which is 11.45% higher than ShallowNet, 8.45% greater than DeepConvNet and 6.10% superior to EEGNet.

*Table 6: Performance comparison of ConTraNet with state-of-the-art methods on 10-class Mendeley Data-sEMG using 5-fold cross validation.*

|  | **Folds** | **EEGNet** (Global) | **DeepConvNet** (Global) | **ShallowNet** (Global) | **ConTraNet** (Global) |
|---|---|---|---|---|---|
|  | 1 | 76.25 | 62.00 | 59.00 | 78.50 |
| **5-Fold CV** | 2 | 76.00 | 72.00 | 69.75 | 79.25 |
|  | 3 | 67.00 | 64.50 | 64.50 | 74.50 |
| **10-class** | 4 | 78.00 | 76.25 | 69.25 | 79.75 |
|  | 5 | 58.00 | 68.75 | 66.00 | 73.75 |
|  | **Avg** | 71.05 | 68.70 | 65.70 | **77.15** |

The results presented in **Table 3**, **Table 4**, **Table 5** and, **Table 6** show that ConTraNet generalizes well on cross-subject as well as within-subject analyses for different HMI paradigms. They also show that ConTraNet can extract and learn distinct features from different HMI domains, however, its counterparts struggle across different paradigms. EEGNet uses depthwise and separable convolutions which may encapsulate well-known EEG feature extraction concepts but struggles to expand towards different domains. Similarly,

DeepConvNet and ShallowNet are tailored to the characteristics of MI-EEG signals which make them suffer big performance loss when used on data of different HMI paradigms.

## 4 Summary and discussion

In this work, we proposed ConTraNet, a single hybrid neural network based on CNN and Transformer architectures for the EEG-HMI and EMG-HMI paradigms. ConTraNet consists of a CNN block and a Transformer block. CNN block introduces the inductive bias in the architecture and learns the local dependencies in the signal whereas the Transformer block learns the global dependencies and pays more attention to the relevant parts of the signal which are significant for correct mapping of the signal into the respective control command and suppresses the non-relevant parts using the attention mechanism. The advantages of using such hybrid architecture are as follows: Firstly, both deep features as well as long range global context are learned from the signal simultaneously, which otherwise would have to be learned by using deep CNN architecture using varying kernel sizes in each layer to increase the receptive field. Secondly, large corpus of data is not required to learn the inductive bias by the transformer as it is now included in the architecture by the CNN block, which otherwise would be required if only transformer-based architecture is used. Thirdly, more attention is focused on the part of the signal which plays a vital role in its classification, unlike just CNN based architectures which use the shared weights and pay uniform attention to the signal in each layer. The attention mechanism in the architecture plays a vital role in arming the model with the abilities to extract and learn the significant and distinct features from different HMI paradigms.

We evaluated and compared the performance of ConTraNet with state-of-the-art algorithms on three publicly available versatile datasets which belong to EEG-HMI and EMG-HMI paradigms: BCI competition IV dataset 2b (MI-EEG), Physionet dataset (MI-EEG) and, Mendeley dataset (sEMG). The results are presented in **Table 3**, **Table 4**, **Table 5** and, **Table 6**. The evaluation results shown in **Table 3**, **Table 5**, and **Table 6** present the generalization capabilities of ConTraNet on aforementioned datasets in cross-subject analysis and its comparison with its counterparts. Results show that in 2-class cross-subject analysis on BCI competition IV dataset 2b, ConTraNet generalized well across subjects compared to reference algorithms. ConTraNet obtained 4.2% improvement in average classification performance across subjects compared to ShallowNet, whereas the improvement is 2.9% and 2.4% compared to DeepConvNet and EEGNet respectively. Similar trend in the performance of ConTraNet can be seen in 2-class, 3-class and, 4-class cross-subject analysis on Physionet dataset, where ConTraNet obtained the average classification accuracy of 83.61% in 2-class decoding task which is 3.09% higher than ShallowNet, whereas it yielded an improvement of 2.67% and 1.8% compared to DeepConvNet and EEGNet respectively. Whereas, in case of 3-class MI-EEG classification, ConTraNet achieved 74.38% average decoding accuracy which is 2.97% higher compared to ShallowNet. However, the improvement is 2.43% and 1.93% compared to EEGNet and DeepConvNet respectively. Similarly, for 4-class MI-EEG task, ConTraNet exhibited an average improvement of 4.2% against the baseline methods. It attained the average classification accuracy of 65.44% which implies an improvement of 6.03% compared to ShallowNet, whereas these numbers are 2.82% and 3.80% compared to DeepConvNet and EEGNet respectively. In the classification 10-classes of sEMG signals,

ConTraNet showed big performance leap compared to the baseline models as it achieved an average classification accuracy of 77.15% which is 11.45% higher than ShallowNet, 8.45% greater than DeepConvNet and 6.10% superior to EEGNet.

However, **Table 4** presents the performance of ConTraNet on BCI competition IV dataset 2b on within-subject analysis. It also shows the effect of transfer learning from global model (trained on data of all subjects except the test subject) on subject specific model (fine-tuned on the training data of test subject). The reported results exhibit the generalization capabilities of ConTraNet in cross-subject as well as within-subject analysis across different HMI paradigms.

These results indicate that the ConTraNet is robust to extract and learn the distinct and wide variety of features from different HMI paradigms. However, the baseline methods EEGNet, DeepConvNet and ShallowNet are design specific for EEG signals and henceforth perform worse on data of different HMI paradigms.

# Acknowledgement

This work is supported by the Ministry of Economics, Innovation, Digitization and Energy of the State of North Rhine-Westphalia and the European Union, grants GE-2-2-023A (REXO) and IT-2-2-023 (VAFES).